\documentclass{article}
\usepackage{amsmath,amsfonts,amsthm}
\usepackage{graphicx}
\usepackage{setspace}
\usepackage[a4paper,left=3cm, right=3cm, top=3cm,bottom=3cm,nomarginpar, includeheadfoot,head=12pt,headsep=18pt]{geometry}
\newtheorem{proposition}{Proposition}
\newtheorem{theorem}{Theorem}
\theoremstyle{remark}
\newtheorem{example}{Example}
\newtheorem{remark}{Remark}
\theoremstyle{definition}
\newtheorem{definition}{Definition}
\title{Improved Frechet bounds and model-free pricing of multi-asset options}
\author{Peter Tankov \\ Centre de Math\'ematiques Appliqu\'ees\\ Ecole Polytechnique 91128 Palaiseau France \\ email: \texttt{peter.tankov@polytechnique.org} }
\date{}
\begin{document}
\maketitle

\begin{abstract}
Improved bounds on the copula of a bivariate random vector are
computed when partial information is available, such as the values of
the copula on a given subset of $[0,1]^2$, or the value of a functional of the copula, monotone with respect to the concordance order. 
 These results are then used to compute model-free bounds on the prices of two-asset options which make use of extra information about the dependence structure, such as the price of another two-asset option.  
\end{abstract}

Key words: copulas, Frechet-Hoeffding bounds, concordance order, basket options.

AMS 2010 subject classification: 60E15, 91G20  

\section{Introduction}
A (two-dimensional) copula is a function $C:[0,1]^2\to [0,1]$ with the following properties: 
\begin{itemize}
\item[i.] Boundary conditions: $C(0,u)=C(u,0)=0$ and $C(1,u)=C(u,1)=u$ for all $u\in[0,1]$. 
\item[ii.]  $C$ is $2$-increasing, \emph{i.e.}~for every $0\leq u_1
  <u_2 \leq 1$ and $0\leq v_1<v_2 \leq 1$, one has 
\begin{align}
C(u_2,v_2)+C(u_1,v_1)-C(u_1,v_2)-C(u_2,v_1)\geq 0. \label{2inc.eq}
\end{align}
\end{itemize}

The classical Frechet-Hoeffding bounds on the distribution function of a two-dimensional random vector, can be expressed in terms of the copula $C$ of this vector:
\begin{align}
W(u,v):=\max(0,u+v-1)\leq C(u,v)\leq \min(u,v):=M(u,v).\label{fh.eq}
\end{align}
In the presence of additional information on the dependence between the components of the vector, these bounds can be narrowed. Nelsen et al. \cite{nelsen.al.01} compute the improved bounds when a measure of association such as Kendall's $\tau$ or Spearman's $\rho$ is given, and the Bertino's family of copulas \cite{fredricks.nelsen.02} yields best possible bounds when the values of the copula on the main diagonal are known. More generally, given a nonempty set of bivariate copulas $\mathcal S$, Nelsen et al. \cite{nelsen.al.04} introduce pointwise best-possible bounds of $\mathcal S$:
$$
{A}(u,v) = \sup\{C(u,v)|C\in \mathcal S\}\quad \text{and} \quad B(u,v) = \inf\{C(u,v)|C\in \mathcal S\}.
$$ 
These bounds are in general not copulas but quasi-copulas, and a fortiori they do not necessarily belong to the set $\mathcal S$. 

In the theoretical part of this paper (section \ref{bounds.sec}), we
first compute the improved Frechet bounds when the values of the
copula on an arbitrary subset of $[0,1]^2$ are given, and provide a
sufficient condition for each bound to be a copula, and therefore, be
the best possible bound. This generalizes the findings of
\cite{nelsen.al.04} on the improved Frechet bounds for copulas with
given diagonal sections. Next, we compute the best-possible bounds
when the value of a real-valued functional of the copula, monotone
with respect to the concordance order and continuous with respect to
pointwise convergence of copulas is given, extending the results of \cite{nelsen.al.01}. 

Since the work of Rapuch and Roncalli \cite{rapuch.roncalli.01} it is known that the prices of most two-asset options, when the marginal laws of the two assets are fixed, become monotone functionals of the copula with respect to the concordance order. The classical Frechet-Hoeffding bounds therefore lead to model-free price estimates for such options \cite{rapuch.roncalli.01,hobson.al.05a,hobson.al.05b}.  

In section \ref{basket.sec}, we obtain a new representation for the
price of a two-asset option, allowing to use a quasi-copula.  This
representation enables us to compute (in section \ref{appli}) the improved model-free estimates of the option's value when the prices of all single-asset options on each of the two assets are known and some extra information about the dependence structure.  This extra information may be, for example, the price of a different two-asset option (for example, zero-strike spread options are often quoted in the market), or the correlation of two assets. This is similar in spirit to a recent work by Kaas et al.~\cite{kaas.al.09} who compute worst-case bounds on the Value at Risk of a portfolio of two assets when the marginals and a measure of association are known. 

\section{Preliminaries}
\label{prelim.sec}
In this section, we recall several useful definitions and results and
fix the notation for the rest of the paper. In the definition of
quasi-copula \cite{genest.al.99}, the $2$-increasing property \eqref{2inc.eq} is replaced by weaker assumptions: 
\begin{definition}
A (two-dimensional) quasi-copula is a function $Q:[0,1]^2\to [0,1]$ with the following properties: 
\begin{itemize}
\item[i.] Boundary conditions: $Q(0,u)=Q(u,0)=0$ and $Q(1,u)=Q(u,1)=u$ for all $u\in[0,1]$. 
\item[ii.] $Q$ is increasing in each argument. 
\item[iii.] Lipschitz property: $|Q(u_2,v_2)-Q(u_1,v_1)|\leq |u_2-u_1|+ |v_2-v_1|$ for all $(u_1,v_1,u_2,v_2)\in[0,1]^4$. 
\end{itemize}
\end{definition}

We denote the set of all copulas on $[0,1]^2$ by $\mathcal C$ and the
set of all quasi-copulas by $\mathcal Q$. The \emph{concordance order}
is the order on $\mathcal Q$ defined by $Q_1 \prec Q_2$ if and only if
$Q_1(u)\leq Q_2(u)$ $\forall u\in \mathbb [0,1]^2$. It is clear that
all quasi-copulas satisfy the Frechet-Hoeffding bounds
\eqref{fh.eq}. Similarly, we say that $Q^n \to Q$ pointwise if
$Q^n(u)\to Q(u)$ $\forall u\in [0,1]^2$. The Lipschitz property
implies that in this case the convergence is uniform in $u$. 

For a copula or a quasi-copula $C$ and a rectangle $R=[u_1,u_2]\times
[v_1,v_2]\subset[0,1]^2$, we define $V_C(R):= C(u_2,v_2)+C(u_1,v_1)-C(u_1,v_2)-C(u_2,v_1)$.

A subset $S\subset [0,1]^2$ is called \emph{increasing} if for all $(a_1,b_1)\in S$ and $(a_2,b_2)\in S$ either $a_1\leq a_2$ and $b_1\leq b_2$ or $a_1\geq a_2$ and $b_1\geq b_2$. It is called \emph{decreasing} if for all $(a_1,b_1)\in S$ and $(a_2,b_2)\in S$ either $a_1\leq a_2$ and $b_1\geq b_2$ or $a_1\geq a_2$ and $b_1\leq b_2$. It is easy to see that for a decreasing set $S$, the set $\bar S:=\{(a,b):(a,1-b)\in S\}$ is increasing. In the same spirit, if $C$ is a copula, the function $\bar C(u,v):=u-C(u,1-v)$ is also a copula and if $Q$ is a quasi-copula, $\bar Q(u,v):=u-Q(u,1-v)$ is also a quasi-copula.

The following well-known result (see e.g. Theorem 3.2.3 in \cite{nelsen2nd}), gives the best-possible bounds of a set of copulas taking a given value at a given point. 
\begin{proposition}\label{single.prop}
Let $C$ be a copula and suppose $C(a,b)=\theta$ with $(a,b)\in[0,1]^2$. Then
\begin{align}
C^{a,b,\theta}_L(u,v)\leq C(u,v)\leq C^{a,b,\theta}_U(u,v),\quad (u,v)\in [0,1]^2,\label{1ptbound.eq}
\end{align}
where
\begin{align*}
C^{a,b,\theta}_U &= \min(u,v,\theta + (u-a)^+ + (v-b)^+)\\ \text{and} \quad C^{a,b,\theta}_L &= \max(0,u+v-1,\theta - (a-u)^+ - (b-v)^+)
\end{align*}
are copulas satisfying $C^{a,b,\theta}_U(a,b)=C^{a,b,\theta}_L(a,b)=\theta$.
\end{proposition}
\begin{remark}
A careful examination of the proof or Theorem 3.2.3 in \cite{nelsen2nd} reveals that \eqref{1ptbound.eq} also holds if $C$ is a quasi-copula satisfying $C(a,b)=\theta$.
\end{remark}

To close this section, we recall a well-known fact on distribution functions. Given a one-dimensional distribution function $F(x)$ we define its generalized inverse by 
$$
F^{-1}(u) = \inf\{x\in \mathbb R: F(x)\geq u\}, \quad u\in(0,1],
$$
with the convention $\inf \emptyset = +\infty$. If the couple $(X,Y)$ has copula $C$ then $(X,Y)$ has the same law as $(F_X^{-1}(U),F_Y^{-1}(V))$, where $(U,V)$ are random variables with distribution function $C$. 
\section{Constrained Frechet bounds}\label{bounds.sec}
Let $S$ be a compact subset of $[0,1]^2$ and $Q$ be a quasi-copula. We denote by $\mathcal C_S$ the set of all copulas $C'$ such that $C'(a,b) = Q(a,b)$ for all $(a,b) \in S$, 
and by $\mathcal Q_S$ the set of all quasi-copulas $Q'$ such that $Q'(a,b) = Q(a,b)$ for all $(a,b) \in S$. Define
\begin{align}
A^{S,Q}(u,v)&:=\min(u,v,\min_{(a,b)\in S}\{Q(a,b)+(u-a)^+ + (v-b)^+\}) \label{asq}\\
B^{S,Q}(u,v)&:=\max(0,u+v-1,\max_{(a,b)\in S}\{Q(a,b)-(a-u)^+ - (b-v)^+\})
\end{align}
The following theorem establishes that $A^{S,Q}$ and $B^{S,Q}$ are best-possible bounds of the set $\mathcal Q_S$. This means that they are also bounds of the set $\mathcal C_S$, but not in general best possible. The second part of the theorem gives a sufficient condition under which $A^{S,Q}$ or $B^{S,Q}$ is a copula, and therefore a best possible bound of $\mathcal C_S$. As a by-product of the second part, we obtain an example of copula which coincides with a given quasi-copula on a given increasing or decreasing set. 

\begin{theorem}${}$
\label{inc.thm}
\begin{itemize}
\item[i.] $A^{S,Q}$ and $B^{S,Q}$ are quasi-copulas satisfying 
$$
B^{S,Q}(u,v) \leq Q'(u,v) \leq A^{S,Q}(u,v)\quad \forall (u,v)\in [0,1]^2.
$$
for every $Q'\in \mathcal Q_S$ and
\begin{align}
A^{S,Q}(a,b)=B^{S,Q}(a,b) = Q(a,b)\label{match}
\end{align}
for all $(a,b)\in S$.
\item[ii.] If the set $S$ is increasing then $B^{S,Q}$ is a copula; if the set $S$ is decreasing then $A^{S,Q}$ is a copula.
\end{itemize}
\end{theorem}
\noindent\emph{The proof of this theorem can be found in the Appendix.}

\begin{example}\label{quasicop.ex}
This example, similar to example 2.1 in \cite{nelsen.al.04} shows that if $S$ is increasing, $A^{S,Q}$ may not always be a copula. Let $S = \left\{\left(\frac{1}{3},\frac{1}{3}\right),\left(\frac{2}{3},\frac{2}{3}\right)\right\}$ and $Q=W$. Then $A^{S,Q}\left(\frac{1}{3},\frac{1}{3}\right)=0$, and $A^{S,Q}\left(\frac{2}{3},\frac{2}{3}\right)=A^{S,Q}\left(\frac{1}{3},\frac{2}{3}\right)=A^{S,Q}\left(\frac{2}{3},\frac{1}{3}\right)=\frac{1}{3}$, so that the $A^{S,Q}$-volume of the rectangle $\left[\frac{1}{3},\frac{2}{3}\right]^2$ is equal to $-\frac{1}{3}$. Similarly, if $S$ is decreasing, $B^{S,Q}$ is not always a copula. 
\end{example}

Let $\rho:\mathcal Q\to \mathbb R$ be a mapping, continuous with
respect to pointwise convergence of copulas and nondecreasing with respect to the concordance order on $\mathcal Q$. We are interested in computing pointwise best possible bounds of the sets $\mathcal C^r := \{C\in \mathcal C: \rho(C)=r\}$ and $\mathcal Q^r := \{Q\in \mathcal Q: \rho(Q)=r\}$. We denote
\begin{align*}
A^r(u,v)&:= \max\{C(u,v)|C\in \mathcal C^r\}\quad \text{and}\quad B^r(u,v):= \min\{C(u,v)|C\in \mathcal C^r\}\\
\tilde A^r(u,v)&:= \max\{Q(u,v)|Q\in \mathcal Q^r\}\quad \text{and}\quad \tilde B^r(u,v):= \min\{Q(u,v)|Q\in \mathcal Q^r\}
\end{align*}
for $(u,v)\in[0,1]^2$. 

For $(a,b)\in [0,1]^2$ and $\theta\in I_{a,b}:=[W(a,b), M(a,b)]$, we define
$$
\rho_+(a,b,\theta):=\rho(C^{a,b,\theta}_U),\quad \rho_-(a,b,\theta):=\rho(C^{a,b,\theta}_L).
$$
For fixed $a,b$, the mappings $\theta\mapsto \rho_+(a,b,\theta)$ and $\theta\mapsto \rho_-(a,b,\theta)$ are nondecreasing and continuous, and we define the corresponding inverse mappings by 
\begin{align*}
r\mapsto \rho^{-1}_-(a,b,r)&:=\max\{\theta\in I_{a,b}:\rho_-(a,b,\theta)=r\}\\
r\mapsto \rho^{-1}_+(a,b,r)&:=\min\{\theta\in I_{a,b}: \rho_+(a,b,\theta)=r\}, 
\end{align*}
for all $r$ such that the corresponding set over which the maximum or minimum is taken is nonempty. 

\begin{theorem}\label{thm2}
Let $r\in[\rho(W),\rho(M)]$. The bounds $A^r, \tilde A^r$ and $B^r, \tilde B^r$ are given by
\begin{align}
A^r(u,v) = \tilde A^r(u,v) &= \left\{\begin{aligned}&\rho^{-1}_-(u,v,r)\quad &&\text{if}\ r\in [\rho(W),\rho_-(u,v,M(u,v))]\\ & M(u,v)\quad &&\text{otherwise}\end{aligned}\right.\\
B^r(u,v)= \tilde B^r(u,v)&= \left\{\begin{aligned}&\rho^{-1}_+(u,v,r)\quad &&\text{if}\ r\in [\rho_+(u,v,W(u,v)),\rho(M)]\\ & W(u,v)\quad &&\text{otherwise}\end{aligned}\right.
\end{align}
\end{theorem}
\noindent\emph{The proof of this theorem can be found in the Appendix.}
\begin{remark}
This result generalizes theorems 2 and 4 in \cite{nelsen.al.01}, which treat the cases when $\rho$ is the Kendall's $\tau$ and the Spearman's $\rho$. In these two cases, $A^r$ and $B^r$ are copulas. However, in general, this may not be the case. Let $(a_1,b_1)\in[0,1]^2$, $(a_2,b_2)\in[0,1]^2$, $W(a_1,b_1)\leq \theta_1 \leq M(a_1,b_1)$, $W(a_2,b_2)\leq \theta_2 \leq M(a_2,b_2)$ and define
$$
\rho(C) = (C(a_1,b_1)-\theta_1)^+ + (C(a_2,b_2)-\theta_2)^+.
$$ 
An easy computation shows that 
$$
A^0(u,v) = \min(u,v,\theta_1+(u-a_1)^++(v-b_1)^+, \theta_2+(u-a_2)^++(v-b_2)^+ ),
$$
that is, we obtain the copula $A^{S,Q}$ of Equation \eqref{asq} with $S=\{(a_1,b_1), (a_2,b_2)\}$ and $Q$ such that $Q(a_1,b_1)=\theta_1$ and $Q(a_2,b_2)=\theta_2$. Then, example \eqref{quasicop.ex} shows that $A^0$ is not always a copula. 
\end{remark}

\section{Copula based pricing of multi-asset options}\label{basket.sec}
We consider the problem of pricing a European-style option whose
pay-off depends on the values of two random variables $X$ and
$Y$. These random variables can represent the terminal values of two
assets (in the context of equity options) or some other risk factors
which influence the value of the option, such as the default dates of
two defaultable bonds. 

We assume that the law of $X$ and $Y$ under the historical probability $\mathbb P$ is unknown, or is very hard to estimate, so that all information comes from the prices of traded options on these assets. 

%Let the discounted pay-off function be denoted by $f(x,y)$. 
Under the standard assumption of absence of arbitrage opportunities in
the market, the option pricing theory implies that there exists a
risk-neutral probability $\mathbb Q$ such that the option price is given by
the discounted expectation of its pay-off under $\mathbb Q$.  
%$$
%\pi =E^{\mathbb Q}[f(X,Y)].
%$$
In practice $\mathbb Q$ is not known, and only some incomplete information on
it can be deduced from the prices of traded options on $X$ and $Y$. 

We assume that these traded options include single-asset options
allowing to reconstruct the cumulative distribution functions $F_X$
and $F_Y$ of $X$ and $Y$. For example, if $X$ is the price of an asset
at time $T$ and call options on this asset with prices
$P_X(K):=E^{\mathbb Q}[e^{-rT}(X-K)^+]$, are available (where $r$ is the interest rate and
$K$ is the strike price), the distribution function can
be reconstructed as $F_X(K) = 1-e^{rT}\frac{\partial P_X(K)}{\partial
  K}$. Similarly, if $X$ is the default date of a defaultable bond,
the distribution function may be reconstructed from the prices of
credit default swaps on this bond with different maturities. 

Let the discounted pay-off function of a two-asset option be denoted
by $f(x,y)$.  Its price then becomes a function of the copula $C$ of $X$ and $Y$:
\begin{align}
 \pi(C)  &= E^{\mathbb Q}[f(X,Y)]= \int_0^\infty \int_0^\infty f(x,y)dC(F_X(x),F_Y(y))\notag\\ &= \int_0^1 \int_0^1 f(F_X^{-1}(u),F_Y^{-1}(v))dC(u,v)
. \label{pricecop.eq}
\end{align}

It is known \cite{muller.scarsini.00,tchen.80} that for every
2-increasing function $f$ such that the integral in
\eqref{pricecop.eq} exists, the mapping $C\mapsto \pi(C)$ is
nondecreasing with respect to the concordance order of
copulas. Therefore, if the pay-off function $f$ is $2$-increasing, and
if we know that the copula $C$ of $X$ and $Y$ satisfies $B\prec C
\prec A$ for two copulas $A$ and $B$, the option price satisfies
$\pi(B)\leq \pi(C) \leq \pi(A)$.  For example, if no additional
information on the joint law of $X$ and $Y$ is available, the standard
Frechet bounds lead to
$$
\int_0^1 \int_0^1 f(F_X^{-1}(u),F_Y^{-1}(v))dW(u,v)\leq \pi(C)\leq\int_0^1 \int_0^1 f(F_X^{-1}(u),F_Y^{-1}(v))dM(u,v).
$$
Since the support of $dM$ is the diagonal $v=u$ and that of $dW$ is
the diagonal $v=1-u$, these bounds are further simplified to 
\begin{align*}
\int_0^1 f(F_X^{-1}(1-u),F_Y^{-1}(u))du \leq \pi(C) \leq \int_0^1 f(F_X^{-1}(u),F_Y^{-1}(u))du.
\end{align*}

However, if $A$ and $B$ are quasi-copulas, this method no longer applies because the integral in \eqref{pricecop.eq} may not be well defined. 
 The following result provides an alternative representation for $\pi(C)$ which can be used for quasi-copulas, and establishes other useful properties of this mapping. We recall \cite[Section 4.5]{kingman66} that for a 2-increasing function $f$ on $[0,\infty)^2$ which is left-continuous in both arguments, there exists a unique positive measure $\mu$ on $[0,\infty)^2$ such that 
\begin{align}
\mu([x_1,x_2)\times[y_1,y_2)) = f(x_1,y_1)+f(x_2,y_2)-f(x_1,y_2)-f(x_2,y_1).\label{induced}
\end{align}   
\begin{proposition}\label{pricecop.prop}
Assume that $f$ is 2-increasing, left-continuous in each of its arguments, and let the marginal laws of $X$ and $Y$ satisfy
$$
E[|f(X,0)|+|f(0,X)|+|f(Y,0)| + |f(0,Y)| + |f(X,X)|+|f(Y,Y)|]<\infty.
$$
Then, $E[|f(X,Y)|]<\infty$ and the mapping $C\mapsto \pi(C)$ is
well-defined for all $C$, continuous with respect to pointwise
convergence of copulas and satisfies
\begin{align}
\pi(C) = &-f(0,0) + E[f(X,0)] + E[f(0,Y)] \notag\\&+ \int_0^\infty\int_0^\infty \mu(dx\times dy)(1-F_X(x)-F_Y(y)+C(F_X(x),F_Y(Y))),\label{pricecop2.eq} 
\end{align}
where $\mu$ is the positive measure on $[0,\infty)^2$ induced by $f$.
\end{proposition}
\noindent\emph{The proof of this proposition can be found in the
  Appendix.}
\begin{remark}
Expression \eqref{pricecop2.eq} can be alternatively written as 
\begin{align*}
\pi(C) = -f(0,0) + E[f(X,0)] + E[f(0,Y)] \notag+
\int_0^\infty\int_0^\infty \mu(dx\times dy)\overline C(\overline
F_X(x),\overline F_Y(Y)), 
\end{align*}
where $\overline C$ is the survival copula defined by
$$
\overline C(u,v) = u+v - 1 + C(1-u,1-v),
$$
and $\overline F_X$ and $\overline F_Y$ are survival functions of $X$
and $Y$. 
\end{remark}
Table \ref{payoffs.tab} gives several examples of 2-asset options whose pay-offs are 2-increasing (or 2-decreasing, meaning that $-f$ is 2-increasing) continuous functions. These are mainly taken from \cite{rapuch.roncalli.01}. For all these pay-offs, the integral with respect to $\mu$ in formula \eqref{pricecop2.eq} reduces to a one-dimensional integral.  Another important example is the function $f(X,Y)=XY$ which is also 2-increasing, which means that for fixed marginal distributions, the linear correlation coefficient
 $$\rho(X,Y) = \frac{E[XY]-E[X]E[Y]}{(\text{Var}\,X\text{Var}\,Y)^{\frac{1}{2}}}$$ 
 is nondecreasing with respect to the concordance order of copulas. The corresponding measure $\mu$ is the Lebesgue measure on $[0,\infty)^2$. 

\begin{table}
\begin{tabular}{p{0.35\textwidth}|p{0.16\textwidth}|l}
\parbox{\textwidth}{Option type\\ and $f(X,Y)$} & \centerline{increasing?} & $\int_0^\infty \int_0^\infty \mu(dx\times dy) G(x,y)$ \\\hline && \\
\parbox{\textwidth}{Basket option,\\ $(\alpha X + \beta Y - K)^+$} & $+$ if $\alpha\beta>0$, $-$ if $\alpha\beta<0$ & $\text{sgn}\, (\alpha \beta) \int\limits_{z: \frac{z}{\alpha}\geq 0, \frac{K-z}{\beta}\geq 0} G\left(\frac{z}{\alpha},\frac{K-z}{\beta}\right)dz.$\\&&\\
\parbox{\textwidth}{Call on the minimum\\ $(\min(X,Y)-K)^+$} & \centerline{$+$} & $\int_K^\infty G(x,x)dx$\\&&\\
\parbox{\textwidth}{Put on the minimum\\ $(K-\min(X,Y))^+$} & \centerline{$+$} &$\int_0^K G(x,x)dx$\\&&\\ 
\parbox{\textwidth}{Call on the maximum\\ $(\max(X,Y)-K)^+$} & \centerline{$-$}  &$-\int_K^\infty G(x,x)dx$\\&&\\
\parbox{\textwidth}{Put on the maximum\\ $(K-\max(X,Y))^+$} &\centerline{$-$} &$-\int_0^K G(x,x)dx$\\ &&\\
\parbox{\textwidth}{Worst-off call\\ $\min((X-K_1)^+ ,(Y-K_2)^+ )$} & \centerline{$+$}& $ \int_0^\infty G(z+K_1,z+K_2)dz$\\&&\\
\parbox{\textwidth}{Worst-off put\\ $\min((K_1-X)^+ ,(K_2-Y)^+ )$} &  \centerline{$+$} & $ \int_0^{\min(K_1,K_2)} G(K_1-z,K_2-z)dz$\\&&\\
\parbox{\textwidth}{Best-off call\\ $\max((X-K_1)^+ ,(Y-K_2)^+ )$} & \centerline{$-$} & $ -\int_0^\infty G(z+K_1,z+K_2)dz$\\&&\\
\parbox{\textwidth}{Best-off put\\ $\max((K_1-X)^+ ,(K_2-Y)^+ )$} & \centerline{$-$} & $ -\int_0^{\min(K_1,K_2)} G(K_1-z,K_2-z)dz$
\end{tabular}
\caption{Common 2-asset option pay-off functions, and the representation of integrals with respect to the corresponding measure $\mu$. The plus sign indicates that the pay-off function is $2$-increasing and the minus that it is $2$-decreasing.}
\label{payoffs.tab}
\end{table}

\section{Application: model-free bounds on option prices}
\label{appli}
In this section, we derive model-free bounds on the prices of
two-asset options whose pay-off function satisfies the assumptions of
Proposition \ref{pricecop.prop} when extra information about the
dependence of $X$ and $Y$ is given. We give four examples
corresponding to different kinds of extra information and different
option pay-offs.

\begin{example}[The case when prices of digital basket options are
  known]
\begin{figure}
\centerline{\includegraphics[width=0.6\textwidth]{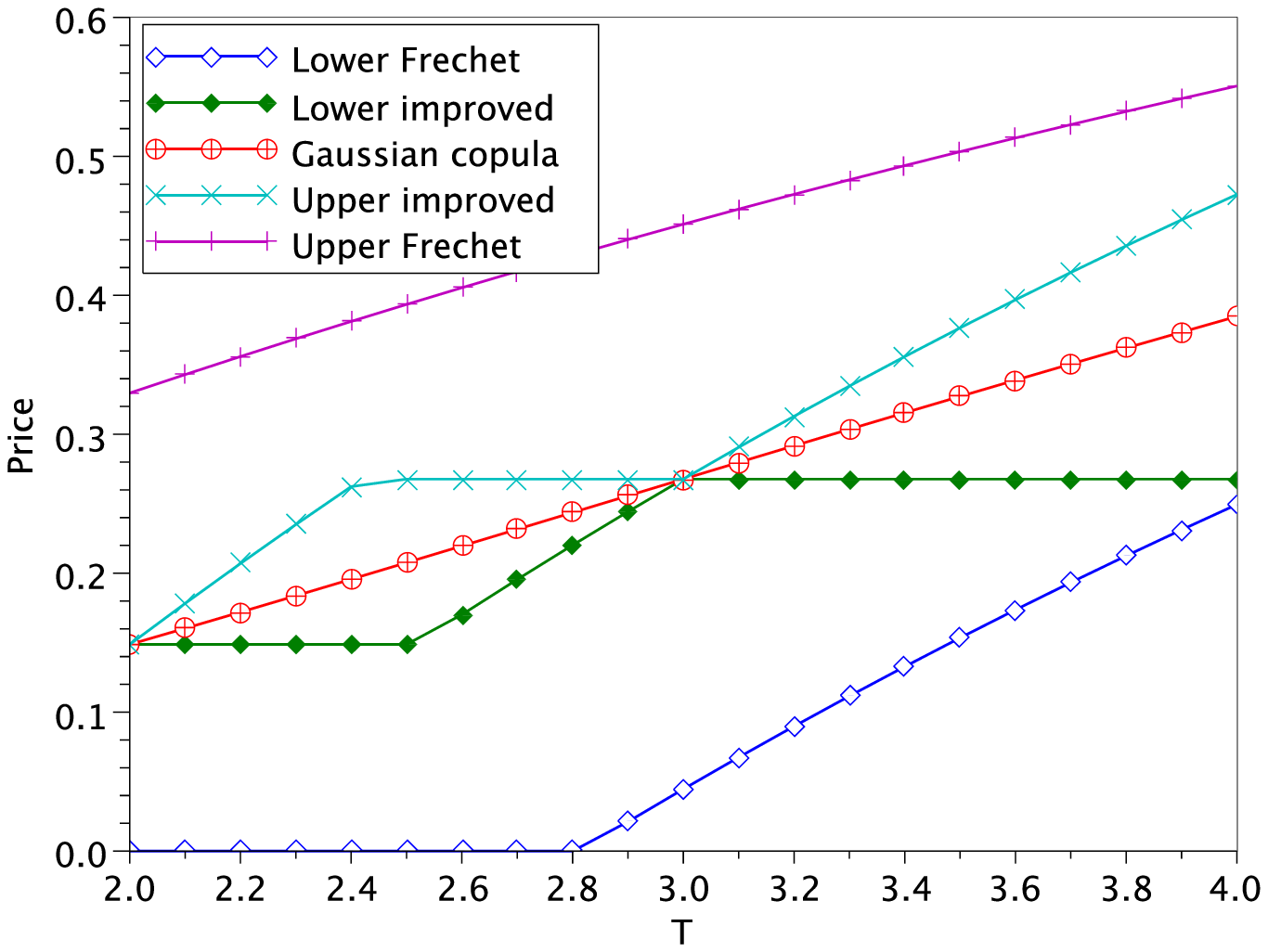}\hspace*{-0.8cm}\includegraphics[width=0.6\textwidth]{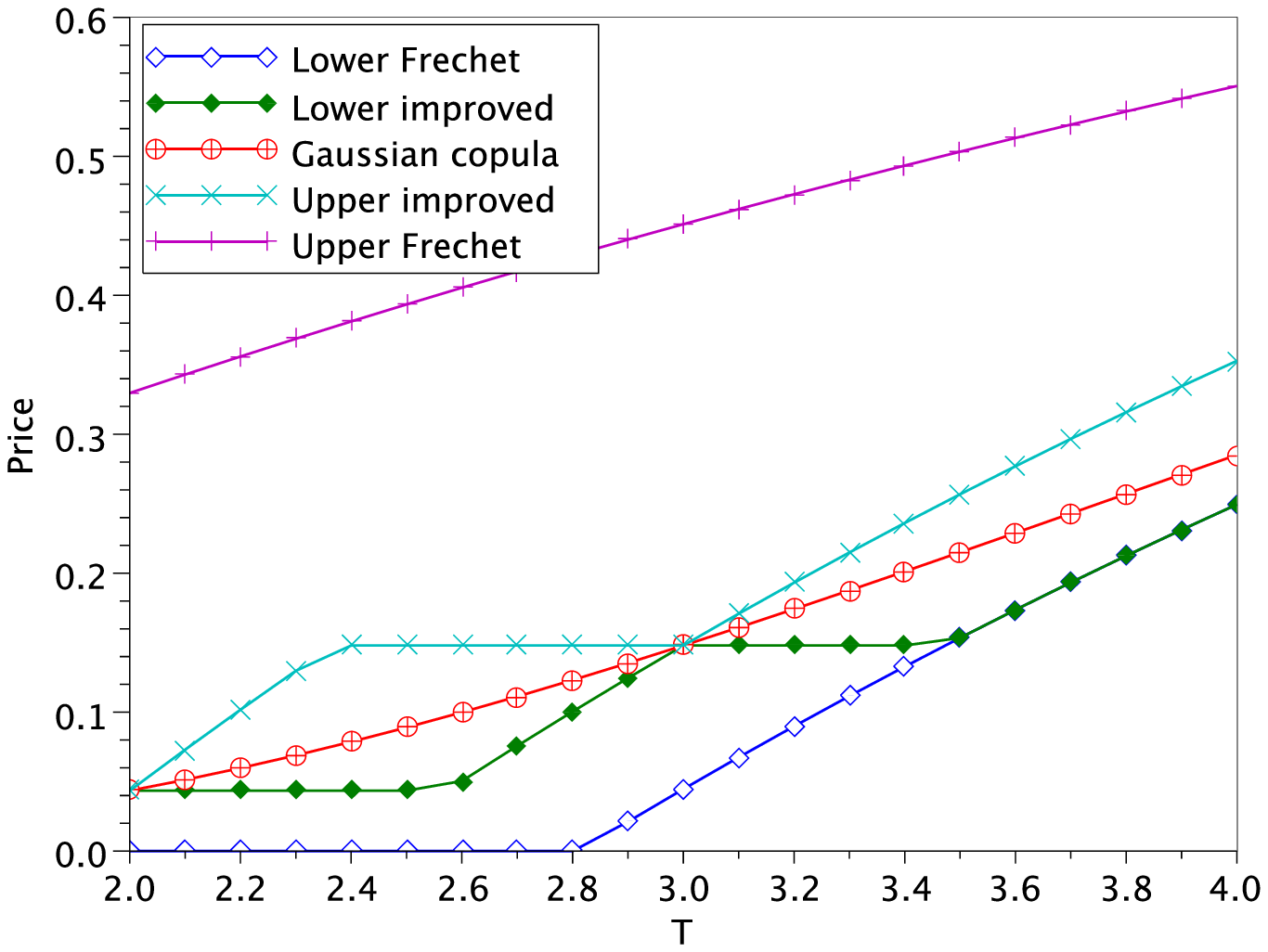}}
\caption{Improved bound on the prices of ``second to default'' option as function of
  time to maturity $T$, when the prices of options with 2 and 3 years
  to maturity are known and equal to the price in the ``Gaussian copula''
  model with given correlation $\rho$. Left: $\rho=0$. Right: $\rho=-0.7$.}
\label{default.fig}
\end{figure}
As our first example, we consider an application to credit risk
modeling assuming that $X$ and $Y$ represent the times of default of
two corporate bonds. In this context, an important problem is the
pricing of the so called ``first to default'' option with pay-off at
maturity $T$ given by $H^1_T = 1_{(X\wedge Y)\leq T}$ or the ``second to
default'' option with pay-off $H^2_T = 1_{(X\vee Y)\leq T} $. The
price of each of these options is directly related to the value of the
copula $C$ of $X$ and $Y$ at the point $(F_X(T),F_Y(T))$:
\begin{align*}
E^{\mathbb Q}[H^1_T] &= 1 - P[X>T, Y> T] = F_X(T) + F_Y(T) -
C(F_X(T),F_Y(T))\\
E^{\mathbb Q}[H^2_T] &= P[X\leq T, Y\leq  T] = C(F_X(T),F_Y(T)).
\end{align*}
In view of the above, we concentrate on the ``second to default''
options.  From the prices $P_k = E[H^2_{T_k}]$ of these options with
maturities $T_1,\dots,T_n$, one can recover the values of the copula
$C$ of $X$ and $Y$ on the increasing set
$(F_X(T_k),F_Y(T_k))_{k=1,\dots,n}$. Therefore, by Theorem
\ref{inc.thm}, the copula $C$ of $X$ and $Y$ satisfies
$$
B(u,v)\leq C(u,v) \leq A(u,v)\quad \forall (u,v)\in[0,1]^2
$$
with 
\begin{align*}
A(u,v) &= \min(u,v,\min_{k=1,\dots,n}\{P_k+(u-F_X(T_k))^+ + (v-F_Y(T_k))^+\})\\
B(u,v) &= \max(0,u+v-1,\max_{k=1,\dots,n}\{P_k-(F_X(T_k)-u)^+ - (F_Y(T_k)-y)^+\}).
\end{align*}
and the price of any 2-asset option whose pay-off function $f(x,y)$ satisfies the assumption of Proposition \ref{pricecop.prop} admits the bounds 
$$
\pi(B) \leq \pi(C) \leq \pi(A).
$$
Since, by Theorem \ref{inc.thm}, $B$ is a copula, the lower bound is sharp, while the upper bound may not necessarily be sharp. 

As an illustration, we have computed the upper and lower improved
bounds for the prices of ``second to default'' options with different
times to maturity. We assume that
the marginal laws of $X$ and $Y$ are exponential with parameters $\lambda_X
= 0.2$ and $\lambda_Y = 0.3$ respectively, and that the prices of the
``second to default''
options with $2$ and $3$ years to maturity are known. In this example,
these two prices are computed assuming that $X$ and $Y$ have Gaussian
copula with correlation $\rho$ (the Gaussian copula is the industry
standard). Figure \ref{default.fig} shows the prices of the ``second to default''
options as function of time to maturity for two different values of $\rho$, along with the price in the
``Gaussian copula'' model and the standard Frechet bounds (without any information about dependence). 
\end{example}

\begin{figure}
\centerline{\includegraphics[width=0.6\textwidth]{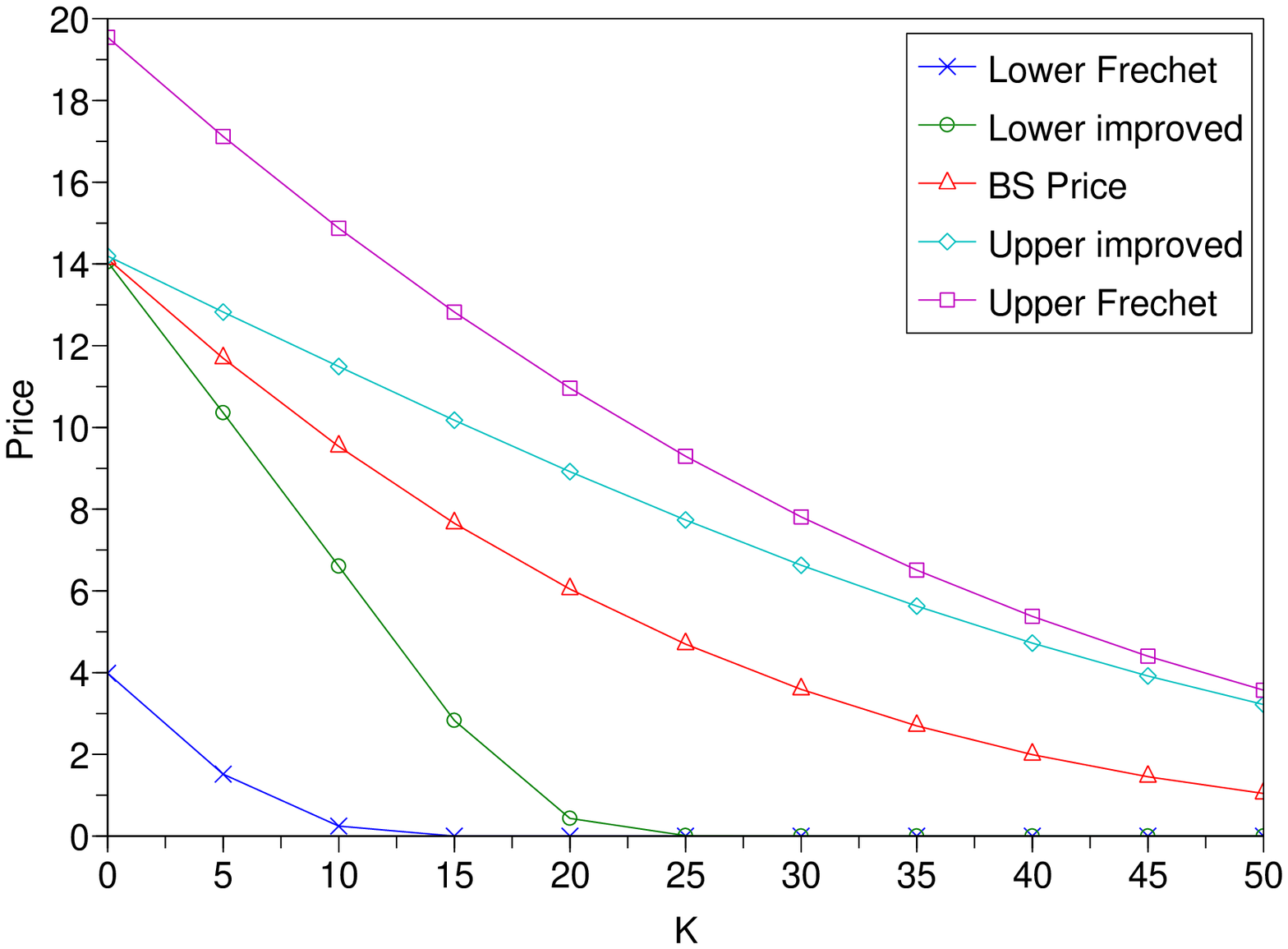}\hspace*{-0.8cm}\includegraphics[width=0.6\textwidth]{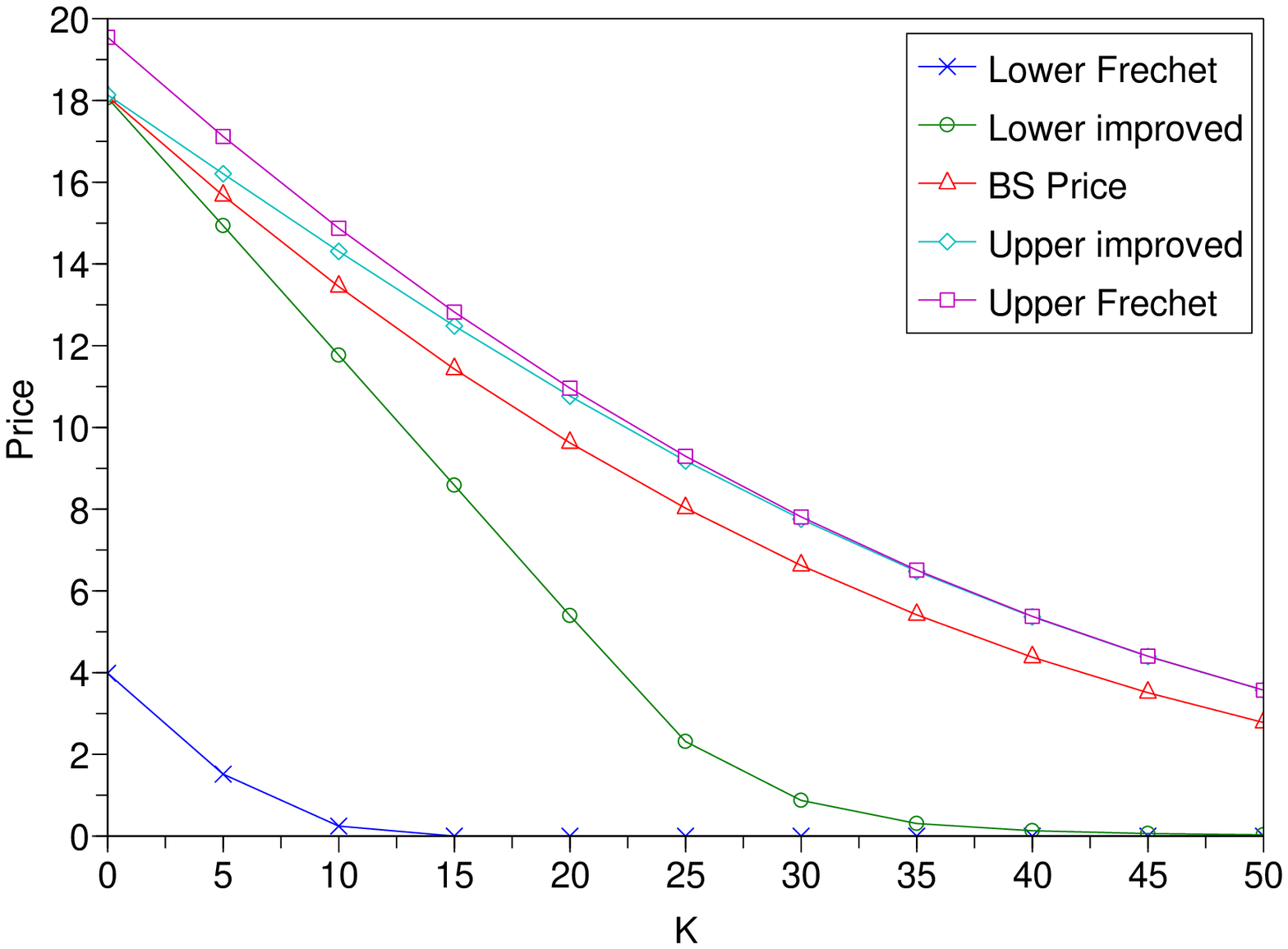}}
\caption{Improved bound on the spread option price as function of
  strike $K$, when the prices of all options on maximum are known and
  equal to the price in the Black Scholes
  model with given correlation $\rho$. Left: $\rho=0$. Right: $\rho=-0.7$.}
\label{spread.fig}
\end{figure}

\begin{example}[The case when prices of all options on the maximum of two assets are known]\label{spread.ex}
The knowledge of prices of call or put options on the maximum or the
minimum of $X$ and $Y$, for all strikes, allows to recover (by
differentiation) the values of the distribution function $F(K,K)$ for
$K\geq 0$, or, equivalently, the values of the copula $C$ on the
increasing set $((F_X(K),F_Y(K)),K\geq 0)$. Therefore, similarly to
the previous example, the copula $C$ of $X$ and $Y$ satisfies
$$
B(u,v)\leq C(u,v)\leq A(u,v)\quad \forall (u,v)\in[0,1]^2,
$$
where 
\begin{align}
A(u,v) &= \min(u,v,\min_{K\geq 0}\{F(K,K)+(u-F_X(K))^+ + (v-F_Y(K))^+\})\label{aex}\\
B(u,v) &= \max(0,u+v-1,\max_{K\geq 0}\{F(K,K)-(F_X(K)-u)^+ - (F_Y(K)-y)^+\}).\label{bex}
\end{align}

To illustrate this method, we have computed the improved upper and
lower bounds for the spread option with pay-off at date $T=1$ given by
$f(X_T,Y_T) = (X_T-Y_T-K)^+$. To fix the marginal laws of $X$ and $Y$,
we assume that $X_t = X_0\exp(\sigma_x W^x_t - \frac{\sigma_x^2
  t}{2})$ and $Y_t = Y_0\exp(\sigma_y W^y_t - \frac{\sigma_y^2
  t}{2})$, where $\sigma_x = 0.2$, $\sigma_y = 0.3$, $X_0=Y_0 = 100$
and $W^y$ and $W^x$ are standard Brownian motions.  We further assume
that the prices of all options on the maximum of $X$ and $Y$ are equal
to the corresponding prices in a model where $W_T^y$ and $W_T^x$ are
jointly Gaussian with correlation $\rho$. 

Figure \ref{spread.fig}
plots the improved bounds on the spread option price as function of
the strike $K$ for two different values of the correlation $\rho$,
along with the Black-Scholes price and the standard Frechet bounds.
For the numerical computation of the bounds, we have taken a discrete set of 400
strikes in \eqref{aex} and \eqref{bex} and used numerical integration
to evaluate \eqref{pricecop2.eq}, which reduces to a one-dimensional
integral in this case. 
 \end{example}
\begin{example}[The case when a single option price is known]\label{singleprice.ex}
Assume now that the extra information about the dependence structure of $X$ and $Y$ is the expectation of a function $f_0$ which satisfies the assumptions of Proposition \ref{pricecop.prop}: $\rho(C) := E^Q[f_0(X,Y)] = r$. In this case, the price of a 2-asset option whose pay-off $f(x,y)$ satisfies the assumptions of Proposition \ref{pricecop.prop} admits the bounds $\pi(B^r) \leq \pi(C) \leq \pi(A^r)$ with $A^r$ and $B^r$ given by Theorem \ref{thm2}. Although $A^r$ and $B^r$ are best-possible bounds of the set of copulas satisfying $\rho(C)=r$, if they are not copulas themselves, the bounds on the option price may not be best possible. 

For the actual computation of $A^r$ and $B^r$ we reduce the expressions for $\rho^+(a,b,\theta)$ and $\rho^-(a,b,\theta)$ to one-dimensional integrals using the results in \cite[section 3.2.3]{nelsen2nd}: 
\begin{align*}
&\rho^+(a,b,\theta) = \int_0^\theta f_0(F_X^{-1}(u),F_Y^{-1}(u))du + \int_\theta^a f_0(F_X^{-1}(u),F_Y^{-1}(u+b-\theta))du\notag\\&\qquad+\int_a^{a+b-\theta} f_0(F_X^{-1}(u),F_Y^{-1}(u+\theta-a))du + \int_{a+b-\theta}^1 f_0(F_X^{-1}(u),F_Y^{-1}(u))du
\end{align*}
\begin{align*}
&\rho^-(a,b,\theta) \\&= \int_0^{a-\theta} f_0(F_X^{-1}(u),F_Y^{-1}(1-u))du + \int_{a-\theta}^a f_0(F_X^{-1}(u),F_Y^{-1}(a+b-\theta-u))du\notag\\&\quad+\int_a^{1-b+\theta} f_0(F_X^{-1}(u),F_Y^{-1}(1+\theta-u))du + \int_{1-b+\theta}^1 f_0(F_X^{-1}(u),F_Y^{-1}(1-u))du
\end{align*}

As the first illustration of this approach, we have computed the improved bounds on the price of the call option on the maximum of two assets, with pay-off at date $T=1$ given by $f(X_T,Y_T) = (\max(X_T,Y_T)-K)^+$, assuming that the price of the zero-strike spread option, with pay-off $f_0(X_T,Y_T) = (X_T-Y_T)^+$, is known (these options are indeed often quoted in the market).  The marginal laws of $X$ and $Y$ are the same as in example \ref{spread.ex}, and  we further assume that the price of the zero-strike spread option is equal to the corresponding price in a model where $W_T^y$ and $W_T^x$ are jointly Gaussian with correlation $\rho$.

Figure \ref{max.fig} plots the improved bounds as function of the
strike $K$ for two different values of the correlation $\rho$, along
with the Black-Scholes price and the standard Frechet bounds. Since we
now have much less information on the dependence of $X_T$ and $Y_T$
than in example \ref{spread.ex}, the improved bounds are not as narrow
as in that example. Still, when the spread option price is close to
one of its extreme values, such as, for example in the right graph of
figure \ref{max.fig}, where we have taken $\rho = -0.7$, the improved
bounds lead to a considerable narrowing of the price interval. In the
numerical example, $\rho^+$ and $\rho^-$ were evaluated by numerical
integration, their inverses were then computed by bissection, and a
further numerical integration was performed to evaluate the bounds.
\begin{figure}
\centerline{\includegraphics[width=0.6\textwidth]{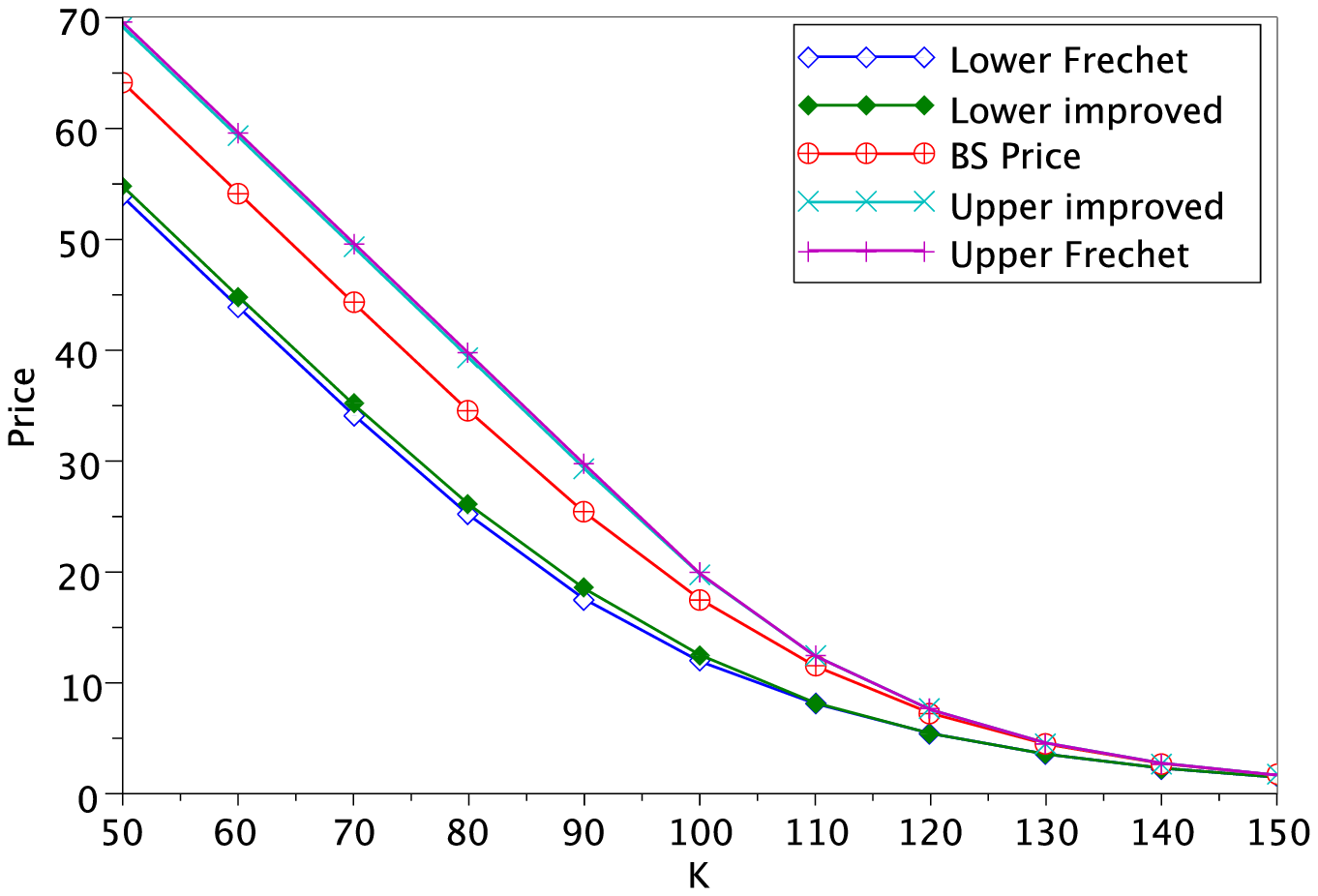}\hspace*{-0.8cm}\includegraphics[width=0.6\textwidth]{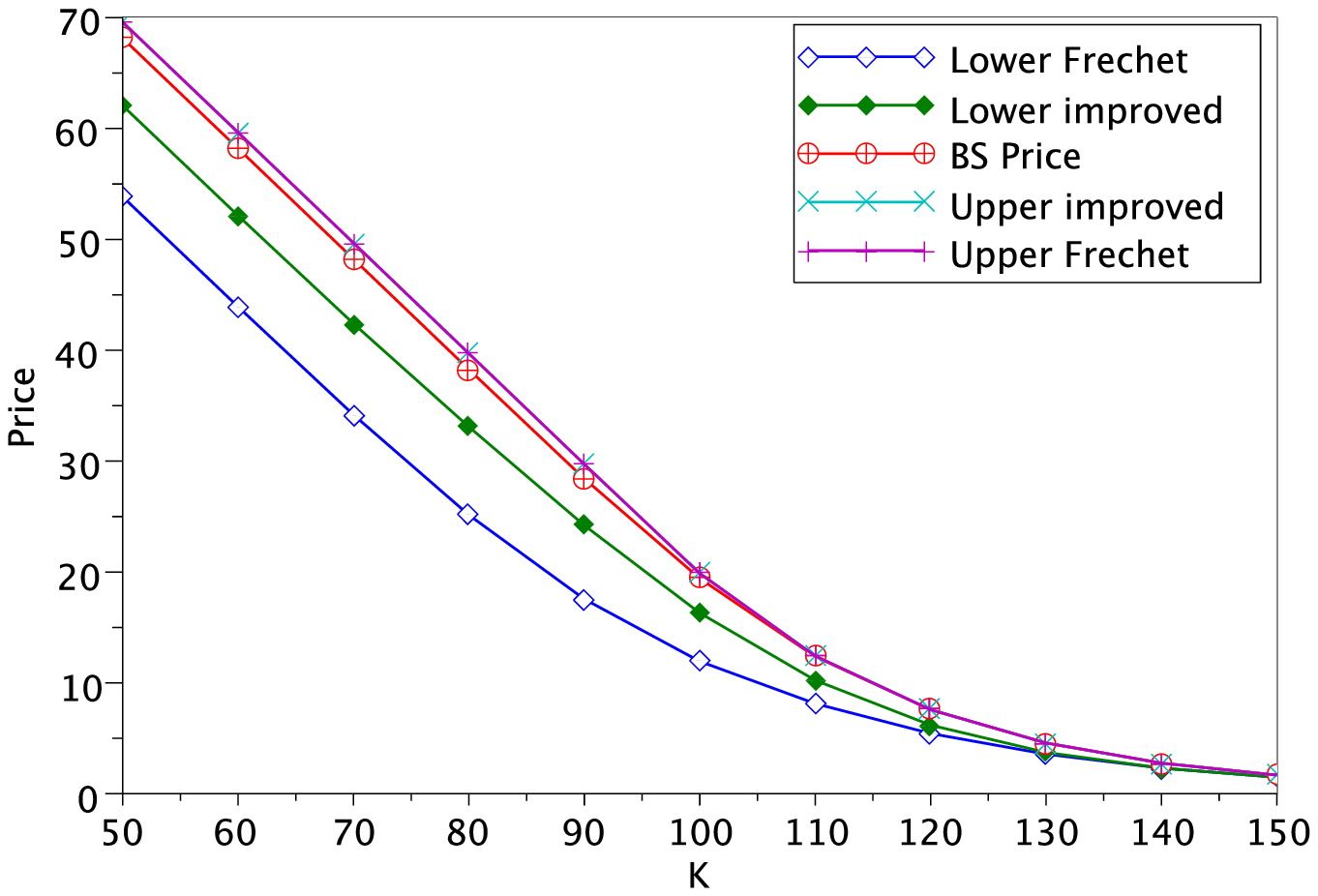}}
\caption{Improved bounds on the price of the option on the maximum of
  two assets as function of strike $K$ when the price of the zero-strike
  spread option is known and equal to the price in the Black-Scholes
  model with given correlation $\rho$. Left: $\rho=0$. Right: $\rho=-0.7$.}
\label{max.fig}
\end{figure}
\end{example}

\begin{example}[The case when the linear correlation of log-returns is
  known]
Very often, the option trader does not know the full two-dimensional
distribution of $X$ and $Y$ under $\mathbb Q$ but has a strong view about the
risk-neutral correlation of log-returns
$$
\rho_0 = \frac{E[\log X \log Y] - E [\log X]\, E [\log Y]}{(\text{Var}(\log X)\, 
  \text{Var}(\log Y))^{\frac{1}{2}}}. 
$$
In this case, one can obtain bounds on the prices of two-asset options
in the same way as in Example \ref{singleprice.ex}, 
using the function $f_0(x,y) = \log x \, \log y$, which is
$2$-increasing. Figure \ref{rhospread.fig} plots the bounds on the
price of a zero-strike spread option with pay-off $f(X_T,Y_T) =
(X_T-Y_T)^+$ when the correlation of log-returns is known, for
different correlation values. As we have already observed in Example \ref{singleprice.ex}, these bounds
are most useful for extreme correlation scenarios, and yield little
additional information when the correlation is close to $0$. 
\begin{figure}
\centerline{\includegraphics[width=0.6\textwidth]{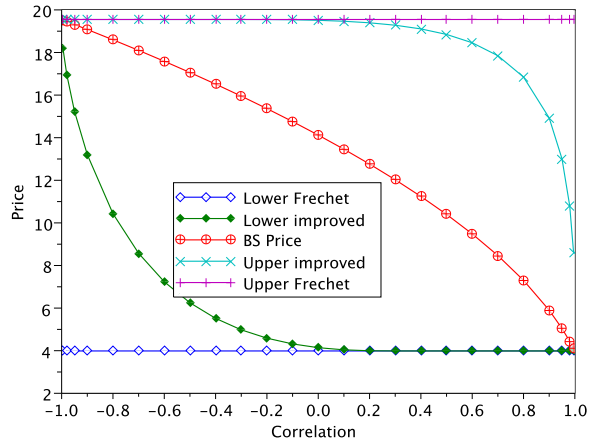}}
\caption{Improved bounds on the price of the zero-strike spread option
when the correlation of log-returns $\rho$ is known, as a function of
$\rho$. For comparison, the price in the Black Scholes model
with the given value of correlation is shown as BS Price.}
\label{rhospread.fig}
\end{figure} 
\end{example}
\section*{Acknowledgement}

This research is supported by the Chair Financial Risks of the Risk Foundation sponsored by Soci\'et\'e G\'en\'erale, the Chair Derivatives of the Future sponsored by the F\'ed\'eration Bancaire
Fran\c caise, and the Chair Finance and Sustainable Development sponsored by EDF and Calyon.

The author would like to thank K. Jean-Alphonse for numerous discussions which helped to shape the paper and for doing preliminary simulations for this research during his internship at Ecole Polytechnique. 
Thanks are also due to the anonymous referee whose
comments helped to improve the presentation of the paper.

\section*{Appendix: the proofs}

\begin{proof}[Proof of Theorem \ref{inc.thm}.]
First, observe that $A$ can be obtained from $B$ by a simple transformation:
$$
A^{S,Q}(u,v) = u-B^{\bar S,\bar Q}(u,1-v) = \overline{B^{\bar S,\bar Q}}(u,v),
$$
where the bar notation was introduced in section \ref{prelim.sec}. It is therefore sufficient to prove only the statements involving $B^{S,Q}$. 

\paragraph{Part i.} Let us first check that $B^{S,Q}$ is a quasi-copula. The boundary conditions follow from the Frechet bounds for $Q$. The fact that $B^{S,Q}$ is increasing in each argument is obvious, and the Lipschitz property follows because for a family of functions $(f_i)_{i\in I}$ which are Lipschitz with constant $1$, we have 
$$
\max_i f_i(y) \leq |x-y| + \max_i f_i(x) \quad \text{and}\quad \max_i f_i(x) \leq |x-y| + \max_i f_i(y),
$$
which implies that $\max_i f_i(x)$ is Lipschitz with the same constant. By proposition \ref{single.prop} and the remark after it,  $C_L^{a,b,Q(a,b)}(u,v) \leq Q'(u,v)$ for all $(u,v)\in[0,1]^2$, $(a,b)\in S$ and $Q' \in \mathcal Q_S$. Since $B^{S,Q}$ is the upper bound of $C_L^{a,b,Q(a,b)}(u,v)$ over $(a,b)\in S$, we have that $B^{S,Q}(u,v) \leq Q'(u,v)$.  

Let us now check the property \eqref{match}. Take $(a',b')\in S$. From the Frechet lower bound for $Q$, we get:
$$
B^{S,Q}(a',b') = \max_{(a,b)\in S} \{Q(a,b)-(a-a')^+ -(b-b')^+\}. 
$$
For every $(a,b)\in S$, using the Lipschitz property of $Q$ and the fact that it is increasing in each argument, we get that 
$$
Q(a,b) - (a-a')^+ -(b-b')^+ \leq Q(a',b') 
$$
Therefore, the $\max$ is attained for $(a,b)=(a',b')$. 

\paragraph{Part ii.} Let $S$ be an increasing set. By adding to this set the points $(0,0)$ and $(1,1)$, we may with no loss of generality simplify the definition of $B^{S,Q}$: 
$$
B^{S,Q}(u,v):=\max_{(a,b)\in S}\{Q(a,b)-(a-u)^+ - (b-v)^+\}
$$
Given that $B^{S,Q}$ is a quasi-copula, we only need to prove property \eqref{2inc.eq}. 

Since $B^{S,Q}$ is Lipschitz continuous, for every $\varepsilon>0$, one can find a finite increasing set $S_\varepsilon$ such that $\sup_{(u,v)\in[0,1]^2}|B^{S_{\varepsilon},Q}(u,v)-B^{S,Q}(u,v)|\leq \varepsilon$. Therefore, it is enough to prove property \eqref{2inc.eq} for a set $S_n = \{(a_i,b_i)\}_{i=1}^n$, where we suppose without loss of generality that $a_i\leq a_{i+1}$ and $b_i\leq b_{i+1}$ for $i=1,\dots, n-1$.

The proof will be done by induction. For $n=1$, property \eqref{2inc.eq} is straightforward. Assume that it holds for $S_n$ and let $a_{n+1}\geq a_n$, $b_{n+1}\geq b_n$ and $S_{n+1}:= S_n \cup \{(a_{n+1},b_{n+1})\}$. To simplify notation, we write $B_n:=B^{S_n,Q}$, $B_{n+1}:=B^{S_{n+1},Q}$ and $Q_{n+1}:=Q(a_{n+1},b_{n+1})$. For convenience, we subdivide the domain $[0,1]^2$ onto four sets $A,B,C$ and $D$ as shown in Figure \ref{4sets.fig}. 

To prove that $B_{n+1}$ is 2-increasing, we must show that for every rectangle $R\subset [0,1]^2$, $V_{B_{n+1}}(R)\geq 0$. However, since $V_B$ is additive over rectangles, it is sufficient to consider only the cases $R\subseteq A$, $R\subseteq B$, $R\subseteq C$ and $R\subseteq D$. By construction, on $A$, the function $B_{n+1}$ only depends on the coordinate $u$, and therefore, $V_{B_{n+1}}(R)= 0$ for every rectangle $R\subseteq A$. Similarly, $V_{B_{n+1}}(R)= 0$ for $R\subseteq B$ because $B_{n+1}$ is constant on $B$ and $V_{B_{n+1}}(R)= 0$ for $R\subseteq C$ because $B_{n+1}$ only depends on the coordinate $v$ on $C$. It remains to consider the case $R\subseteq D$.  

Let $R = [u_1,u_2]\times [v_1,v_2]\subseteq D$. We must show that 
\begin{align*}
V_{B_{n+1}}(R) & = \max(B_n(u_1,v_1), Q_{n+1}-(a_{n+1}-u_1) - (b_{n+1}-v_1))\\
& + \max(B_n(u_2,v_2), Q_{n+1}-(a_{n+1}-u_2) - (b_{n+1}-v_2))\\
& - \max(B_n(u_1,v_2), Q_{n+1}-(a_{n+1}-u_1) - (b_{n+1}-v_2))\\
& - \max(B_n(u_2,v_1), Q_{n+1}-(a_{n+1}-u_2) - (b_{n+1}-v_1))\geq 0
\end{align*}
We consider separately three cases. 
\begin{itemize}
\item[--] If $B_n(u_1,v_2)\geq Q_{n+1}-(a_{n+1}-u_1) - (b_{n+1}-v_2)$ and $B_n(u_2,v_1)\geq Q_{n+1}-(a_{n+1}-u_2) - (b_{n+1}-v_1)$ then $V_{B_{n+1}}(R)\geq B_n(u_1,v_1) + B_n(u_2,v_2) - B_n(u_1,v_2)-B_n(u_2,v_1)\geq 0$ by induction hypothesis. 
\item[--] Assume $B_n(u_1,v_2)\leq Q_{n+1}-(a_{n+1}-u_1) - (b_{n+1}-v_2)$. Then, by the Lipschitz property of $B_n$, necessarily $B_n(u_2,v_2)\leq Q_{n+1}-(a_{n+1}-u_2) - (b_{n+1}-v_2)$, and therefore, by the Lipschitz property of $B_{n+1}$, $V_{B_{n+1}}(R)= u_2 - u_1 +  B_n(u_1,v_1) - B_n(u_2,v_1)\geq 0$. 
\item[---] The remaining case, when $B_n(u_1,v_2)\geq Q_{n+1}-(a_{n+1}-u_1) - (b_{n+1}-v_2)$ and $B_n(u_2,v_1)\leq Q_{n+1}-(a_{n+1}-u_2) - (b_{n+1}-v_1)$, is treated similarly to the second one. 
\end{itemize}

\begin{figure}
\centerline{\begin{picture}(120,120)(-20,-10)
\put(0,0){\line(0,1){100}}
\put(0,0){\line(1,0){100}}
\put(100,0){\line(0,1){100}}
\put(0,100){\line(1,0){100}}
\put(0,70){\line(1,0){100}}
\put(70,0){\line(0,1){100}}
\put(35,35){\makebox{D}}
\put(35,80){\makebox{A}}
\put(83,80){\makebox{B}}
\put(83,35){\makebox{C}}
\put(-10,-10){\makebox{0}}
\put(-20,70){\makebox{$b_{n+1}$}}
\put(65,-10){\makebox{$a_{n+1}$}}
\put(103,-10){\makebox{1}}
\put(-10,100){\makebox{1}}
\end{picture}}
\caption{Illustration for the proof of Theorem \ref{inc.thm}, part ii.}
\label{4sets.fig}
\end{figure}
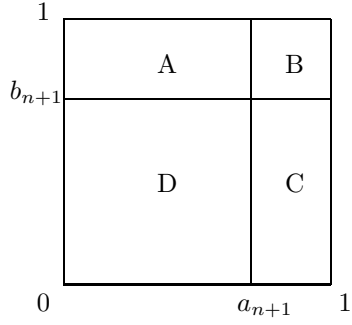
\end{proof}

\begin{proof}[Proof of Theorem \ref{thm2}.]
We give the proof for the bound $\tilde A^r(u,v)$. Since the proof is only based on Proposition \ref{single.prop} which holds in the same form both for copulas and for quasi-copulas, $A^r$ coincides with $\tilde A^r$. The proof for lower bounds $\tilde B^r$ and $B^r$ is similar. 

Assume $r\in [\rho(W),\rho_-(u,v,M(u,v))]$. Then, since $\theta\mapsto \rho_-(u,v,\theta)$ is increasing and continuous, $\rho(C^{u,v,\rho_-^{-1}(u,v,r)}_L) = r$ and therefore, $A^r(u,v)\geq \rho_-^{-1}(u,v,r)$. On the other hand,
$$
\{\rho(Q)|Q(u,v) = \theta\} \subseteq [\rho_-(u,v,\theta),\rho_+(u,v,\theta)].
$$
By definition of $\rho_-^{-1}$, for all $\theta>\rho^{-1}_-(u,v,r)$, $\rho_-(u,v,\theta)>r$ and therefore for every $Q\in \mathcal Q$ such that $Q(u,v)>\rho^{-1}_-(u,v,r)$, $\rho(Q)>r$. Therefore, $A^r(u,v)\leq \rho_-^{-1}(u,v,r)$. 

Assume now that $r>\rho_-(u,v,M(u,v))$ and let $C^w:=(1-w)C_L^{u,v,M(u,v)} + wM(u,v)$. Then $\rho(C^0)<r$, $\rho(C^1)\geq r$ (by assumption of the theorem) and since $\rho$ is continuous, there exists $w\in[0,1]$ such that $\rho(C^w)=r$. Since $C^w(u,v)=M(u,v)$ for all $w$, this proves that $A^r(u,v)\geq M(u,v)$. On the other hand, clearly $A^r(u,v)\leq M(u,v)$ (Frechet bound). 
\end{proof}

\begin{proof}[Proof of Proposition \ref{pricecop.prop}.]
Since $f$ is 2-increasing, $V_f([0,x]\times[0,y]) = f(x,y) + f(0,0) - f(x,0)-f(0,y)$ is increasing in $x$ and $y$, and therefore
\begin{align*}
&|f(x,y)|=|V_f([0,x]\times[0,y])-f(0,0)+f(x,0)+f(0,y)|\\
&\quad \leq |f(0,0)| + |f(x,0)| + |f(0,y)| + |V_f([0,x]^2)| + |V_f([0,y]^2)| \\
&\quad \leq C\{|f(0,0)| + |f(0,x)| + |f(x,0)|+|f(0,y)| + |f(y,0)| + |f(x,x)| + |f(y,y)|\},
\end{align*}
for some $C>0$, which implies $E[|f(X,Y)|]<\infty$. 

%Since $f$ is 2-increasing and left-continuous, it induces a positive measure $\mu$ on $[0,\infty)^2$ such that for all $x,y\geq 0$,
%$$
%f(x,y) = \mu([0,x)\times [0,y)) - f(0,0) + f(x,0)+f(0,y). 
%$$
Let $p(dx\times dy)$ be the law of $(X,Y)$. By Fubini's theorem and \eqref{induced} we then get
\begin{align*}
\pi(C) &= E[f(X,Y)] = -f(0,0) + E[f(X,0)] + E[f(0,Y)] \\&+ \int_0^\infty\int_0^\infty\int_0^\infty\int_0^\infty 1_{x'<x}1_{y'<y} \mu(dx'\times dy') p(dx\times dy)\\ & =  -f(0,0) + E[f(X,0)] + E[f(0,Y)] \\ &+ \int_0^\infty\int_0^\infty  \mu(dx'\times dy') P[X>x',Y>y'] \\ &= -f(0,0) + E[f(X,0)] + E[f(0,Y)] \\&+ \int_0^\infty\int_0^\infty \mu(dx\times dy)(1-F_X(x)-F_Y(y)+C(F_X(x),F_Y(Y)))
\end{align*}
In the last integral, the integrand is positive and bounded from above
by the function $1-F_X(x)-F_Y(y)+\min(F_X(x),F_Y(y))$, which
corresponds to the copula of complete dependence and is integrable by
the first part of the proposition. Therefore, the dominated
convergence theorem implies that $\pi(C)$ is continuous with respect
to pointwise convergence of copulas. 
\end{proof}

\end{document}